\documentclass[superscriptaddress,twocolumn,showpacs,a4paper,
amssymb,amsmath,nobibnotes,aps,prd,longbibliography,
showkeys,
nofootinbib,notitlepage]{revtex4-1}
\usepackage{graphicx}
\usepackage{float}
\usepackage{epsf}
\usepackage{bm}
\usepackage{amsmath}
\usepackage{amsfonts}
\usepackage{amssymb}
\usepackage{epstopdf}
\usepackage{natbib}
\usepackage{color}%
\providecommand{\U}[1]{\protect\rule{.1in}{.1in}}
\newcommand{\be}{\begin{equation}}
\newcommand{\ee}{\end{equation}}

\newcommand{\mincir}{\raise
-3.truept\hbox{\rlap{\hbox{$\sim$}}\raise4.truept\hbox{$<$}\ }}
\newcommand{\magcir}{\raise
-3.truept\hbox{\rlap{\hbox{$\sim$}}\raise4.truept\hbox{$>$}\ }}

\newtheorem{remark}{Remark}[section]
\usepackage{xcolor}
\usepackage{hyperref}
\hypersetup{
    colorlinks=true, 
    linkcolor=blue,
    citecolor=magenta}

\newcommand{\R}{\mathbb{R}}
\begin{document}

\title{On the perturbed Friedmann equations in Newtonian Gauge}

\author{Jaume de Haro}
\email{jaime.haro@upc.edu}
\affiliation{Departament de Matem\`atiques, Universitat Polit\`ecnica de Catalunya, Diagonal 647, 08028 Barcelona, Spain}

\author{Emilio Elizalde}
\email{elizalde@ice.csic.es}
\affiliation{Institute of Space Science, ICE/CSIC and IEEC,
Campus UAB, C/Can Magrans, s/n, 08193 Bellaterra, Barcelona, Spain}

\author{Supriya Pan}
\email{supriya.maths@presiuniv.ac.in}
\affiliation{Department of Mathematics, Presidency University, 86/1 College Street,  Kolkata 700073, India}
\affiliation{Institute of Systems Science, Durban University of Technology, PO Box 1334, Durban 4000, Republic of South Africa}

\begin{abstract}
Based on the Newtonian mechanics, in this article, we present a heuristic derivation of the Friedmann equations, providing an intuitive foundation for these fundamental relations in cosmology. Additionally, using the first law of thermodynamics and Euler's equation, we derive a set of equations that, at linear order, coincide with those obtained from the conservation of the stress-energy tensor in General Relativity. This approach not only highlights the consistency between Newtonian and relativistic frameworks in certain limits but also serves as a pedagogical bridge, offering insights into the physical principles underlying the dynamics of the universe.
\end{abstract}

\vspace{0.5cm}


\maketitle

\section{Introduction}

\noindent In the initial phase of our study, we delve into the Friedmann equations, originally formulated by Alexander Friedmann in the 1920s as the solutions to Einstein's field equations of General Relativity (GR). These equations provide a foundational framework for understanding the large-scale evolution of the universe, incorporating the assumptions of homogeneity and isotropy while accounting for curvature and energy content.\\

\noindent Friedmann's pioneering work \cite{Friedman:1922kd,Friedmann:1924bb}, which described the solutions for both expanding and contracting universes, went largely unnoticed for years \cite{Klimchitskaya:2022ryr}. Even Einstein, at first, rejected Friedmann's findings, arguing that they were inconsistent with GR. However, Einstein later acknowledged their validity, though he initially remained unconvinced about the possibility of an expanding universe. It was only with the subsequent discovery of the Hubble-Lema\^{i}tre law, which established a connection between observational cosmology and fundamental physics, that Friedmann's contributions gained widespread recognition. This synthesis of GR with thermodynamics laid the groundwork for a deeper understanding of cosmic evolution.\\

\noindent Building on this foundation, we explore an alternative derivation of the Friedmann equations, inspired by earlier works such as those by McCrea and Milne \cite{1934QJMat...5...73M} and by Callan, Dicke, and Peebles \cite{1965AmJPh..33..105C}. Using Newtonian mechanics, Poisson's equation and the first law of thermodynamics, we present a heuristic approach that, while simpler in conception, aligns with the relativistic results 
at the first order of perturbations.\\

\noindent We also generalize the classical fluid dynamics equations—the continuity and  Euler equations—to ensure their compatibility with GR under first-order perturbations. This involves extending these equations to account for the gravitational effects predicted by GR, providing a unified framework that captures the interplay between matter, spacetime, and perturbations. By reconciling the classical and relativistic approaches, we aim to present a holistic understanding of the dynamics governing the universe's evolution, from large-scale homogeneity to the complexities introduced by perturbations.\\

\noindent The structure of the article is very simple. With section~\ref{sec-historical}
we start with a historical background of Friedmann equations. Later in section~\ref{sec-3} we show the derivation of the Friedmann equations from the Newtonian mechanics. In section~\ref{sec-4-perturbations} we discuss the perturbation equations. In section~\ref{sec-final-equations} we list the key equations for describing the evolution of the universe in the Newtonian approach. Finally, in section~\ref{sec-conclusion} we draw our conclusions. 

\section{Historical note on the Friedmann's Equations}
\label{sec-historical}

\noindent Alexander Friedmann was born in 1888 in St. Petersburg, Russia, where he
remained for much of his short life. In 1910, he obtained his bachelor's degree from the city's State University, and later became a Professor at the  Saint Petersburg Mining Institute. During World War I, Friedmann fought as an aviator of the Russian Imperial army. Upon his return to Petrograd (the new name for St. Petersburg), he finally defended his master's thesis: {\it ``The hydromechanics of a compressible fluid''.} Subsequently, he acquired a keen interest in the mathematics underlying Einstein's GR,  published four years earlier but still quite unknown in Russia. \\

\noindent Friedmann knew Paul Ehrenfest during the five years Ehrenfest spent in St. Petersburg. 
The Ehrenfest archive at the Lorentz Institute in Leiden contains several documents Friedmann sent to Ehrenfest. Late in 1920, Friedmann wrote to him: {\it ``I have been working on the axiomatics of the relativity principle, starting from two propositions: {\it a)} uniform movement continues to be so for all observers in uniform movement; {\it b)} the speed of light is constant (the same for both a static and a moving observer). I have obtained formulas, for a one-dimensional universe, more general than the Lorentz transformations''} \cite{alma991058533603806011}.\\

\noindent In April 1922, he wrote another letter to Ehrenfest, in Russian, saying (translated):
{\it ``I am sending you a brief note on the shape of a possible universe, more general than Einstein's cylindrical and de Sitter's spherical ones. Apart from these two cases, a world also appears whose space has a radius of curvature that varies with time. It seemed to me that such a question might interest you or de Sitter. As soon as I can, I will send you a German translation of this note. And, if you think that the question under consideration is interesting, please be so kind as to endorse it to a scientific journal''.} This work {\it ``On the question of the geometry of a space with curvature''}, dated April 15, 1922, was never published. Actually, Ehrenfest sent the manuscript along with a letter Friedmann had written to Hermann Weyl to the mathematician Jan Schouten, who provided a critical report on Friedmann's analysis.\\

\noindent In the same year 1922, and while all this was going on, Friedmann translated his article into German. He had elaborated it further, changed the title to: {\it ``On the curvature of space''}, and finally decided to send it directly to Zeitschrift f\"{u}r Physik for publication. It was received by the journal on June 29, 1922. In the document, Friedmann showed that the curvature radius of the universe could be a function of an increasing or periodic time. Friedmann himself commented on the results of that article in a book he wrote later: {\it ``The case of a stationary universe comprises only two possibilities, which were previously considered by Einstein and de Sitter. The case of a variable universe admits, on the contrary, a great number of possible situations. For some of them, the curvature radius of the universe increases constantly with time. And there are other situations that correspond to a radius of curvature that changes periodically''}.\\

\noindent Einstein analyzed Friedmann's article rather quickly, as his response was
received by Zeitschrift f\"{u}r Physik on September 18, 1922: {\it ``The results regarding the non-stationary universe contained in the work look
suspicious to me. It happens that the solution given for this case does not satisfy the field equations''}.\\

\noindent Friedmann learned of this response from his friend Yuri Krutkov, who was visiting Berlin. On December 6, Friedmann wrote to Einstein a letter saying:
{\it ``Taking into account that the possible existence of a non-stationary universe is of some interest, I would like to present to you here the calculations that I have made, so that you can verify and critically evaluate them. [Here he  detailed the operations]. If you find that the calculations that I present in this letter are correct, please be so kind as to inform the editors of Zeitschrift f\"{u}r Physik about it. Perhaps in this case you may want to publish a correction to the statement you made, or give me the opportunity to have a portion of this letter published''}.\\

\noindent However, Einstein had left on a six-month trip to the far and middle East, which finished in Spain. He only returned to Berlin in March 1923, but did not react to Friedmann's letter. In May, Krutkov and Einstein met in Leiden, on the occasion of Hendrik Lorentz's retirement as Professor. There, in Ehrenfest's house, Krutkov told Einstein the details contained in Friedmann's letter. Finally, on May 18, 1923, Krutkov wrote to his sister: 
{\it ``I have managed to defeat Einstein in the argument of Friedman's work. Petrograd’s honor is saved!''}. 
Einstein had admitted his mistake and immediately wrote to Zeitschrift f\"{u}r
Physik retracting his previous note: {\it ``In my previous note I criticized Friedmann's work on the curvature of space. However, a letter from Mr. Friedmann, forwarded to me by Mr. Krutkov, has convinced me that my criticism had been based on a mistake made in my own calculations. I now consider that Mr. Friedmann's results are correct and shed new light''}.
The retraction was received in Zeitschrift f\"{u}r Physik on May 31, 1923.
However, Einstein had not been convinced that Friedmann's solutions were of any use. For at least a decade, no one considered Friedmann's work
as a possible model for our Universe. It took Einstein ten more years to finally admit the expansion of the Universe as a physical possibility, despite the astronomical evidence that had been accumulating already. \\

\noindent In retrospect, it is now easy to recognize that the revolution that eventually gave birth to modern cosmology had started in 1912 \cite{Elizalde:2020kpn}, the year in which Henrietta Leavitt published her groundbreaking results on cepheid  variable stars (crucial for the determination of distances) and Vesto Slipher begun his project to obtain the radial speeds of spiral nebulae from their spectra (displaced towards blue or red because of the optical Doppler effect). At the 17th meeting of the American Astronomical Society held in August 1914 at Northwestern University in Evanston, Illinois, Slipher presented results with speeds of 15 spiral nebulae. All but three of the nebulae were moving away at considerable speeds, which were very difficult to reconcile with a static universe model. They shook the very foundations of this hitherto accepted model. And Slipher's presentation was so clear and convincing that they were received by the audience with a standing ovation. Among the public, a young astronomer, Edwin Hubble, was very impressed by those findings, as he would confess towards the end of his life. They oriented his further research, which led to the celebrated Hubble-Lema\^{i}tre law \cite{Elizalde:2018wsg}.

\section{Friedmann Equations from Newtonian mechanics}
\label{sec-3}

The concept of merging cosmology with Newtonian mechanics has a long history, dating back to works such as those by Milne and Peebles \cite{1934QJMat...5...73M,1965AmJPh..33..105C}. These pioneering efforts laid the groundwork for exploring how Newtonian mechanics, traditionally used to describe local systems, could be extended to provide insights into the large-scale dynamics of the universe.  

In this study, we derive the Friedmann equations for a barotropic fluid in a flat Friedmann-Lema\^{i}tre-Robertson-Walker (FLRW) spacetime. Following the approach outlined in
\cite{deHaro:2024dhe} (see also \cite{Minazzoli:2012md}), we employ a Lagrangian formulation to bridge Newtonian mechanics and cosmology. 
We begin by considering the following Lagrangian
\begin{eqnarray}\label{NewtonianLagrangian1}
{L}_N
=\frac{\dot{a}^2}{2}+\frac{4\pi G}{3}a^2\rho_0,
\end{eqnarray}
where ``a'' denotes the scale factor of the FLRW universe, an overhead dot represents the cosmic time derivative and $\rho_0$ is the homogeneous energy density of the universe. 
Using the Euler-Lagrange equation and the first law of thermodynamics, $d(\rho_0 a^3)=ad(\rho_0 a^2)+\rho_0 a^2da \Longrightarrow -3p_0 a^2da= ad(\rho_0 a^2)+\rho_0 a^2da \Longrightarrow d(\rho_0 a^2)=-(3p_0+\rho_0)ada$, 
we find the second Friedmann equation as,
\begin{eqnarray}\label{second_Friedmann}
    \frac{\ddot{a}}{a}=-\frac{4\pi G}{3}(\rho_0+3p_0),
\end{eqnarray}
where $H = \dot{a}/a$, is the Hubble rate of the FLRW universe. 
Now, using some simple algebra, one can derive the first Friedmann equation. This can be done by combining the second Friedmann equation and the  first law of thermodynamics, which is expressed as,
\begin{eqnarray}\label{cons}
\dot{\rho}_0=-3H(\rho_0+p_0).
\end{eqnarray}
In order to do so, we rewrite (\ref{second_Friedmann}) in the following way,
\begin{eqnarray}\label{A}
\frac{\ddot{a}}{a}=-{4\pi G}( p_0+\rho_0)+\frac{8\pi G}{3}\rho_0,\end{eqnarray}
and taking the derivative of $H$ with respect to the cosmic time, one can  easily obtain 
\begin{eqnarray}\label{A1}
\frac{d H^2}{dt}=2H\frac{\ddot{a}}{a}-2H^3.
\end{eqnarray}
Now, inserting (\ref{A}) into it (\ref{A1}) and using (\ref{cons}), we obtain,
\begin{eqnarray}
&&\frac{d}{dt}\left(H^2-
\frac{8\pi G}{3}\rho_0
\right)=-2H\left(H^2-
\frac{8\pi G}{3}\rho_0
\right),
\end{eqnarray}
whose solution is given by
\begin{eqnarray}
H^2-\frac{8\pi G}{3}\rho_0=\frac{C}{a^2},\end{eqnarray}
and by setting the constant of integration $C$ equal to zero, we obtain the first Friedmann equation in the flat FLRW space-time, and taking $C=-k$, where $k$ represents the spatial curvature, we obtain the general case.

Finally, 
it is useful to write the Friedmann equations, in the flat FLRW space-time, in terms of the conformal time $\eta$ which is related to the cosmic time $t$ as  $ad\eta=dt$, as
\begin{eqnarray}
    \frac{a''}{a}=\frac{4\pi G}{3}a^2T_0,\quad\mathcal{H}^2=\frac{8\pi G}{3}a^2\rho_0,\end{eqnarray}
where $T_0=\rho_0-3p_0$, is the trace of the stress-tensor, prime stands for the derivative with respect to the conformal time, and $\mathcal{H} = a'/a$ is the conformal Hubble parameter.

\section{Perturbations}
\label{sec-4-perturbations}

We begin with the following metric in the weak field approximation for a static field $|\Phi_{\rm N}|\ll 1$,
\begin{eqnarray}\label{metric}
ds^2=(1+2\Phi_{\rm N}({\bf x}))dt^2- { (1-2\Phi_{\rm N}({\bf x})) }d{\bf x}^2,
\end{eqnarray}
which is identical with formula (106.3) of \cite{Landau:1975pou} (also, it is present in the Einstein's book titled ``The Meaning of Relativity'' \cite{Einstein1953-EINTMO-8}). 
This, according to the present-day language, is cited as the ``Newtonian gauge''.  Here, ${\bf x}$ denotes the physical coordinate. 
Now, including the expansion of the universe characterized by the scale factor,
and the variability of the Newtonian potential, the metric (\ref{metric}), in co-moving coordinates ${\bf q}$ in which ${\bf x}=a{\bf q}$,  assumes the form below
\begin{eqnarray}
ds^2=(1+2\Phi_{\rm N}({\bf q}, t))dt^2- a^2(t){ (1-2\Phi_{\rm N}({\bf q},t)) }d{\bf q}^2,
\end{eqnarray}
 where ``$a(t)(1-\Phi_{\rm N}({\bf q},t))$'' (labeled as $a_{\rm N}({\bf q},t)$)is 
the perturbed scale factor  and it plays a crucial role in the understanding of the dynamics of the universe. 
Now writing $\rho=\rho_0+\delta \rho$,  $p=p_0+\delta p$, ($\delta$ indicates the mild variation of the respective quantity) 
and express the first Friedmann equation in a spatially flat FLRW universe in the following form: 
\begin{eqnarray} 
    {\mathcal H}((a^2)'-{\mathcal H}a^2)=\frac{8\pi G a^4}{3} \rho_0.
\end{eqnarray}
On the other hand, the Poisson equation  is given by: 
\begin{eqnarray}
    \Delta_{\bf x}\Phi_N=4\pi G \delta\rho,
\end{eqnarray}
which using the co-moving 
co-ordinates 
has the form 
\begin{eqnarray}
    -\frac{1}{3}\Delta_{\bf q}(-2a^2\Phi_N)=\frac{8\pi G a^4}{3}\delta \rho.
\end{eqnarray}
In the perturbed case, since the scale factor is
 $a_{\rm N}=a(1-\Phi_{N})$, therefore, one can have
$a^2_{\rm N}\cong a^2(1-2\Phi_{N})$. To reconcile both equations, and considering the linearity, the first Friedmann equation can be written as
{
\begin{eqnarray}
{\mathcal H}(\partial_{\eta}(a_{\rm N}^2)-{\mathcal H}a_{\rm N}^2)-\frac{1}{3}\Delta_{\bf q}a_{\rm N}^2=\frac{8\pi G a^4}{3}\rho.
\end{eqnarray}}
On the other hand, the second Friedmann equation can be written as 
\begin{eqnarray}
(a^2)''-2{\mathcal H}^2a^2=\frac{8\pi G a^4}{3} T_0.
\end{eqnarray}
For the sake of convenience, we rescale the coordinates $\bar{\bf x}=\sqrt{3}{\bf x}$, and we obtain, 
\begin{eqnarray}
    \Delta_{\bar{\bf x}}\Phi_N=\frac{4\pi G}{3} \delta\rho.
\end{eqnarray}
In the spirit of the first early scalar theories of gravity \cite{Nordstrom1912,Nordstrom1913,Abraham,Abraham1}, the natural generalization of this equation is
\begin{eqnarray}
\partial^2_{t^2}\Phi_{\rm N}-\Delta_{\bar{\bf x}}\Phi_N=-\frac{4\pi G}{3} \delta T.
\end{eqnarray}

\begin{remark}
    Note that in co-ordinates $(t,{\bf x})$, this equation takes the form:    
\begin{eqnarray}
\partial^2_{t^2}\Phi_{\rm N}-c_s^2\Delta_{{\bf x}}\Phi_N=-\frac{4\pi G}{3} \delta T,
\end{eqnarray}
where the sound speed is $c_s=\frac{1}{\sqrt{3}}$ -  characteristic of a relativistic fluid. Consequently, in this framework, the Newtonian potential $\Phi_{\rm N}$ is a sound wave  traveling with velocity 
 $1/\sqrt{3}$, and driven by the source perturbation 
 $\delta T$.
\end{remark}
Then, 
the simplest dynamical equation containing  both equations is:
\begin{eqnarray}
\Box_{\bar{\bf q}} a_{\rm N}^2
-2{\mathcal H}^2 a_{\rm N}^2=\frac{8\pi G a^4}{3} T,
    \end{eqnarray}
    where we have introduced the D'Alembertian 
    $\Box_{\bar{\bf q}}\equiv\partial^2_{\eta^2}-\Delta_{\bar{\bf q}}$.
Therefore, the perturbed Friedmann equations are:
\begin{eqnarray}\label{perturbed-friedmann-equations}
{\mathcal H}(\partial_{\eta}(a_{\rm N}^2)-{\mathcal H}a_{\rm N}^2)-\Delta_{\bar{\bf q}}a_{\rm N}^2=\frac{8\pi G a^4}{3}\rho, 
\nonumber\\
\Box_{\bar{\bf q}}a_{\rm N}^2
    -2{\mathcal H}^2 a_{\rm N}^2=\frac{8\pi G a^4}{3} T.\end{eqnarray}
It is important to realize that these equations coincide with the first order equations in GR. In effect, at the first order, the perturbed equations in GR are \cite{Mukhanov:2005sc}
\begin{eqnarray}
    \Delta_{\bf q}\Phi_{\rm N}-3{\mathcal H}(\partial_{\eta}\Phi_{\rm N}+{\mathcal{H}}\Phi_{\rm N})=4\pi G a^2\delta\rho,\\
\partial_{\eta^2}^2\Phi_{\rm N}+3\mathcal{H}
\partial_{\eta}\Phi_N+
(2\mathcal{H}'+\mathcal{H}^2)\Phi_{\rm N}=4\pi G a^2\delta p, \end{eqnarray}
which,  after combination, lead, up to first order,  to the perturbed Friedmann equations (\ref{perturbed-friedmann-equations}).
Finally, we want to stress that the generalized second Friedmann equation can be obtained from the variation with respect to $\Psi_N\equiv a_{\rm N}^2$ of the Lagrangian
\begin{align}
{\mathcal L}=\frac{1}{2}
\left[{(\partial_{\eta}\Psi_{\rm N})^2}-{|\nabla_{\bar{\bf q}}\Psi_{\rm N}|^2}+2{\mathcal H}^2\Psi_{\rm N}^2\right]+\frac{8\pi Ga^4}{3} T \Psi_{\rm N}.
\end{align}

\subsection{Conservation and Euler's equations}

Once we have obtained the perturbed Friedmann equations, we can find the evolution equations for the energy density and the velocity of the perturbations, which of course, can be obtained from the conservation equation $\nabla_{\mu}T^{\mu}_{\nu}=0$. However, we want to find them from the first principle of thermodynamics and the classical Euler's equation.

First of all, we want to obtain the conservation equation in GR. To do it, we need the following result: {\it 
Let $\varphi_{\eta}: {\R}^3 \rightarrow {\R}^3$ be the flow of a perfect fluid, with $\varphi_0 $ being the identity mapping. We define the vector velocity ${\bf u}(\varphi_{\eta}({\bf q}), \eta) = \frac{d\varphi_{\eta}({\bf q})}{d\eta}$.
Then, we arrive at the crucial result (see for details \cite{Girbau}):}
\begin{eqnarray}
\left[\frac{d}{d\eta}\int_{\varphi_{\eta}(V)} f({\bf q}, \eta) dq_1dq_2dq_3\right]_{\eta=\bar{\eta}} \nonumber\\ =
\int_{\varphi_{\bar{\eta}}(V)}\left( {\partial_{\eta} f}+ \nabla_{\bf q} \cdot (f{\bf u}) \right)_{\eta=\bar{\eta}}
dq_1dq_2dq_3.
\end{eqnarray}
Now, we consider a small co-moving volume $\delta V$, where  we assume that within this volume element, pressure is the same in all the points. 
Then, 
applying this result to the first law of thermodynamics we obtain
\begin{eqnarray}
&&\left[\frac{d}{d\eta}\int_{\varphi_{\eta}(\delta V)} \rho({\bf q}, \eta) dV\right]_{\eta=\bar{\eta}} \nonumber\\
&&=
-p({\bf q}, \eta)
\left[\frac{d}{d\eta}\int_{\varphi_{\eta}(\delta V)} 1 dV\right]_{\eta=\bar{\eta}}, \end{eqnarray}
where the element of volume is $dV=a_{\rm N}^{3}dq_1dq_2dq_3$.
This integral is equivalent to the differential equation
  \begin{eqnarray}
    D_{\eta}\rho+(\rho+p)\left[3\mathcal{H}_{ \rm N}
  +\nabla_{\bf q}.  {\bf u}  \right]=0,
  \end{eqnarray} where we have used the total derivative 
$D_{\eta}\rho=\frac{\partial\rho}{\partial\eta}+{\bf u}.\nabla_{\bf q}\rho$ and 
we have introduced the total Hubble rate $\mathcal{H}_{\rm N}\equiv \frac{D_{\eta}a_{\rm N}}{a_{\rm N}}$.
Retaining only the first order perturbation, one gets:
\begin{align}
\partial_{\eta}\delta\rho+3{\mathcal H}(\delta\rho+\delta p)+(\rho_0+p_0)\left[
  \nabla_{\bf q}.   {\bf u}  -3{\partial_{\eta} \Phi_{\rm N}}
    \right]=0,
  \end{align}  
which coincides with the first order perturbation of the GR, and at zero order one obviously obtains the conservation equation in the flat FLRW space-time.  Finally, we need the equation for the evolution of ${\bf u}$.
Firstly, we recall that for a dust fluid, i.e., $|p|\ll \rho$, taking the line element $ds^2=-dt^2+d{\bf q}^2$ in the Minkowski space-time, the classical Euler's equation can be written as:
\begin{eqnarray}\label{Eulerintegral}
\left[\frac{d}{dt}\int_{\varphi_{t}(V)} 
\rho{\bf v} dV\right]_{t=\bar{t}}=
\int_{\partial \varphi_{\bar t}(V)} {\bf \mathcal T}({\bf n})dS
\nonumber\\-\int_{\varphi_{\bar t}(V)}\rho\nabla_{\bf q} \Phi_{\rm N}
dV,\end{eqnarray}
where $ {\bf \mathcal T}:\R^3\longrightarrow \R^3$ is the stress tensor,
${\bf v}=\frac{d{\bf q}}{dt}$, 
$dS$ is the element of area, and ${\bf n}$ is the external unit vector to the boundary. Taking into account that for a perfect fluid one has,
$ {\bf \mathcal T}({\bf n})=-p{\bf n}$, and from the Gauss theorem, the Euler equation 
can be written as: 
\begin{align}\label{Eulerintegral1}
\left[\frac{d}{dt}\int_{\varphi_{t}(V)} 
\rho{\bf v} dV\right]_{t=\bar{t}}=
-\int_{\varphi_{\bar t}(V)} (\nabla_{\bf q}p+
\rho\nabla_{\bf q} \Phi_{\rm N})
dV,\end{align}
whose 
 differential form is:
\begin{align}\label{Euler}
\partial_t(\rho{\bf v})+ \nabla_{{\bf q}}.(\rho{\bf v}){\bf v}+
\rho{\bf v}.\nabla_{{\bf q}}{\bf v}+\nabla_{{\bf q}} p+\rho \nabla_{{\bf q}} \Phi_{\rm N}=0,
\end{align}
or
\begin{eqnarray}
\partial_t{{\bf v}}+
{\bf v}.\nabla_{{\bf q}}{\bf v}+\frac{1}{\rho}\nabla_{{\bf q}} p+ \nabla_{{\bf q}} \Phi_{\rm N}=0, \end{eqnarray}
where we have used the first law of thermodynamics for a dust fluid or the continuity equation $\partial_t{\rho}+\nabla_{{\bf q}}.(\rho{\bf v})=0$.
Note that the equation (\ref{Euler}) is incompatible with special relativity. For this reason, we will compare it with the conservation law $\nabla_{\mu} T^{\mu}_{\nu}=0$ in the flat FLRW spacetime.
Recall that the energy-stress tensor is :
\begin{eqnarray}
    T^{\mu\nu}=(\rho+p)u^{\mu}u^{\nu}-pg^{\mu\nu},
\end{eqnarray}
and we have the relation
\begin{align}
    \nabla_{\mu}T^{\mu}_{\nu}=\frac{1}{\sqrt{-g}}\partial_{\mu}(\sqrt{-g}T^{\mu}_{\nu})-\frac{1}{2}\partial_{\nu}(g_{\mu\alpha})T^{\mu\alpha}.
\end{align}

For the metric $ds^2=a^2(d\eta^2-d{\bf q}^2)$, we ahve:
\begin{align}
    u^{\mu}=\left(
    \frac{d\eta}{ds}, \frac{d{\bf q}}{ds}\right)=\frac{1}{a\sqrt{1-|{\bf u}|^2}}(1,{\bf u})\cong \frac{1}{a}(1,{\bf u}),
\end{align}
since we assume that the velocity of the fluid is much smaller than the speed of light. Then, $\nabla_{\mu} T^{\mu}_{k}=0$,
which is equivalent to 
$\partial_{\mu}(a^4T^{\mu}_k)=0$, because $\partial_kg_{\mu \nu}=0$, 
leads to:
\begin{eqnarray}
\ \partial_{\eta}((\rho+p){\bf u})+
4{\mathcal H}(\rho+p){\bf u}
+\nabla_{\bf q}.((\rho+p){\bf u}){\bf u}\nonumber\\+
(\rho+p){\bf u}.\nabla_{\bf q} {\bf u}+\nabla_{\bf q} p=0. \end{eqnarray}
Therefore, the Euler equation in an expanding universe is compatible with the special relativity and the expansion of the universe is obtained by replacing the mass density $\rho$ by the heat function per unit volume $(\rho+p)$ and ${\bf v}$ by
$a{\bf  u}$. 
Therefore, taking into account this replacement, when one includes gravity, the integral form of the Euler's equation becomes:
\begin{eqnarray}\label{Eulerintegral2}
\left[\frac{d}{a_{\rm N}d\eta}\int_{\varphi_{\eta}(V)} 
a_{\rm N}(\rho+p){\bf u} dV\right]_{\eta=\bar{\eta}}
\nonumber\\=
-\int_{\varphi_{\bar \eta}(V)} \left(\nabla_{\bf q}p
-\frac{
\rho+p}{a_{\rm N}}
\nabla_{\bf q} a_{\rm N}\right)
dV,\end{eqnarray}
with $dV=a_{\rm N}^{3}dq_1dq_2dq_3$ and where we have made the replacement $\nabla_{\bf q} \Phi_{\rm N}\rightarrow
-\frac{1}{a_{\rm N}}
\nabla_{\bf q} a_{\rm N}$.
In differential form this becomes:
\begin{eqnarray}
   \frac{1}{a_{\rm N}} \partial_{\eta}[ a_{\rm N}^{4} (\rho+p)
    {\bf u}  ]+ {\bf u}.\nabla_{\bf q}[a_{\rm N}^{3} (\rho+p)
    {\bf u}]
    \nonumber\\
    +a_{\rm N}^{3}\left[ (\rho+p)(\nabla_{\bf q}.{\bf u}){\bf u}+
    \nabla_{\bf q}p-
    \frac{
\rho+p}{a_{\rm N}}
\nabla_{\bf q} a_{\rm N}
\right]=0,
\end{eqnarray}
which, using the total derivative and disregarding some second order terms,  can be written as:
\begin{align}
    D_{\eta}\left( (\rho+p)
    {\bf u}\right)
   + (\rho+p)\left[(4{\mathcal H}_{\rm N}
   +\nabla_{\bf q}.{\bf u}){\bf u}-\frac{1}{a_{\rm N}}
\nabla_{\bf q} a_{\rm N}   
\right] \nonumber\\+ \nabla_{\bf q}p=0.
\end{align}
Note that, at the first order of perturbations, the equation becomes:
\begin{eqnarray}
\partial_{\eta}\left( (\rho_0+p_0)
    {\bf u}\right)
    +4{\mathcal H}(\rho_0+p_0) {\bf u}   
     +(\rho_0+p_0)\nabla_{\bf q} \Phi_{\rm N}     \nonumber\\+ \nabla_{\bf q}\delta p
=0, 
\end{eqnarray}
which coincides with the linearization of $\nabla_{\mu}T_k^{\mu}=0$.

\section{The final equations}
\label{sec-final-equations}

In this final section, we  synthesize all the equations derived throughout this work, bringing together the key results to provide a cohesive overview of our findings. By consolidating these equations, we aim to highlight the interconnectedness of the different approaches and methodologies employed, ranging from heuristic derivations of the Friedmann equations to the analysis of perturbations and the integration of classical and relativistic frameworks.

We have seen that
dealing in  Newtonian gauge, considering $|\Phi_{\rm N}|\ll 1$, the metric can be written as
\begin{eqnarray}
ds^2=a^4a_{\rm N}^{-2}d\eta^2- a_{\rm N}^2d{{\bf q}}^2,
\end{eqnarray}
where $a_{\rm N}=a(1-\Phi_{\rm N})$ is the perturbed scale factor.
From  this metric, we have found the following dynamical equations: 
\begin{enumerate}
    \item First Friedmann equation:
\begin{eqnarray}
(2{\mathcal H}{\mathcal H}_{\rm N}
-{\mathcal H}^2)a_{\rm N}^2-\frac{1}{3}\Delta_{\bf q}a_{\rm N}^2=\frac{8\pi G a^4}{3}\rho.
\end{eqnarray}

\item Second Friedmann equation:
\begin{eqnarray}\label{secondfriedmannequation}
    \left( \partial_{\eta^2}^2-\frac{1}{3}\Delta_{\bf q}\right)
    a_{\rm N}^2
    -2{\mathcal H}^2 a_{\rm N}^2=\frac{8\pi G a^4}{3} T.\end{eqnarray}
    \item Conservation equation or the first law of thermodynamics:
\begin{eqnarray}\label{Conservation_final}
D_{\eta}\rho+3\mathcal{H}_{ \rm N}    (\rho+p)\left[1+
  \frac{1}{3\mathcal{H}_{ \rm N}}\nabla_{{\bf q}}.  {{\bf u}}  \right]=0,  
  \end{eqnarray} where $\mathcal{H}_{\rm N}\equiv \frac{D_{\eta}a_{\rm N}}{a_{\rm N}}$ is the total Hubble rate. 
  
\item Euler's equation:

\begin{eqnarray}\label{Euler_final}
 D_{\eta}\left( (\rho+p)
    {{\bf u}}\right)
   + 4{\mathcal H}_{\rm N}   (\rho+p)\Bigg[\left(1
   +\frac{1}{4{\mathcal H}_{\rm N}}
   \nabla_{{\bf q}}.{{\bf u}}\right){{\bf u}}\nonumber\\-\frac{1}{4{\mathcal H}_{\rm N}a_{\rm N}}\nabla_{{\bf q}} a_{\rm  N}\Bigg] +    \nabla_{{\bf q}}p=0.
   \end{eqnarray}

\end{enumerate}
To conclude, 
 it is important to recognize that, in this approach, as in FLRW spacetime, a privileged frame emerges: the co-moving frame, which moves with the fluid filling the universe. In fact, since we are dealing with an evolution problem we  proceed as in the Arnowitt–Deser–Misner (ADM) formalism~\cite{Arnowitt:1962hi}, where
 the Lorentz manifold has a privileged foliation  by space-like surfaces $\Sigma_{\eta}$,
 and co-moving curved coordinates are adopted.  Then, in Newtonian Gauge the metric, namely $\frak{g}$,  has the form: 
\begin{align}\label{metricafinal}
ds^2=a^4a_{\rm N}^{-2}d\eta^2-a_{\rm N}^2\gamma_{ij}dq^idq^j
\Longleftrightarrow
\nonumber\\
ds^2=a^4a_{\rm N}^{-2}d\eta^2-a_{\rm N}^2\gamma(d{\bf q}, d{\bf q}). \end{align}
Dealing with  curved co-ordinates, 
the Laplacian must be replaced by the Laplace-Beltrami operator relative to the stationary metric, $\gamma$, i.e.,  $ \Delta_{\bf q}a_{\rm N}^2\rightarrow\frac{1}{\sqrt{\gamma}}
\partial_i(\sqrt{\gamma}\gamma^{ij}\partial_j a_{\rm N}^2)$, or, in a compact geometrical language, 
$\mbox{div}_{\gamma}(\nabla a_{\rm N}^2)$. Then, the second Friedmann equation will become:
\begin{align}
\partial_{\eta^2}^2 
a_{\rm N}^2-\frac{1}{3}
\mbox{div}_{\gamma}(\nabla a_{\rm N}^2)
-2{\mathcal H}^2a_{\rm N}^2=\frac{8\pi Ga^4}{3}T,
\end{align}
which is covariant under {\it passive} diffeomorfisms in co-moving co-ordinates. 
In addition, in the conservation and Euler's  equations, the divergence of the velocity is given by
$\nabla_{\bf q}.{\bf u}=\frac{1}{\sqrt{\gamma}}
\partial_i(\sqrt{\gamma}u^i)
$ 
and the gradient of the pressure is expressed as
$(\nabla p)^i=\gamma^{ij}\partial_j p$.
 Note that the second Friedmann equation can be improved considering the conformal metric $ds=a_{\rm N}d\bar{s}$, that is,
 \begin{align}
d\bar{s}^2=\bar{g}(d\eta,d\eta)-\gamma(d{\bf q},d{\bf q}),
 \end{align}
with $\bar{g}(d\eta,d\eta)=\bar{g}_{00}d\eta^2=\left(a/a_{\rm N}
\right)^4 d\eta^2$
and $\sqrt{\bar g}=
\left(a/a_{\rm N}
\right)^2$.
Dealing with  this metric,
we  have obtained 
the  geometric form of the  second Friedmann equation,
\begin{align}\label{Friedmann_final}
\frac{1}{a_{\rm N}^2}\left(\mbox{div}_{\bar g}\nabla -\frac{1}{3}
\mbox{div}_{\gamma}\nabla 
-2\bar{g}({\bf H}_{\rm N}, {\bf H}_{\rm N})\right)a_{\rm N}^2
=\frac{8\pi Ga^2_{\rm N}}{3}T,
\end{align}
where  
$\mbox{div}_{\bar g}(\nabla a_{\rm N}^2)=
\frac{1}{\sqrt{\bar g}}\partial_{\eta}\left(\sqrt{\bar g}\bar{g}^{00}\partial_{\eta}
a_{\rm N}^2\right)$, is the Laplace-Beltrami operator relative to the metric $\bar{g}$
applied to $a_{\rm N}^2$
and
$\bar{g}({\bf H}_{\rm N}, {\bf H}_{\rm N})=\bar{g}^{00}
{\mathcal H}_{\rm N}^2$ --where we have introduced the  Hubble vector rate relative to the metric $\bar{g}$,   $  {\bf {H}}_{\rm N}\equiv\bar{g}^{00}{\mathcal H}_{\rm N}\partial_{\eta}=
\bar{g}^{00}
\frac{\partial_{\eta}a_{\rm N}}{a_{\rm N}}\partial_{\eta}$-- is the square of the total Hubble rate.

Another way to write eqn. (\ref{Friedmann_final}) is to introduce the normal time-like vector ${\bf n}=
\bar{g}^{00}\sqrt{\bar g}\partial_{\eta}=
\left(a_{\rm N}/a\right)^2\partial_{\eta}$. Then,
the final form of the second Friedmann equation, which is the main result of this work, goes as follows:
\begin{align}\label{Friedmann_final1}
\frac{1}{a_{\rm N}^2}\left(
\partial^2_{{\bf n}^2}
-\frac{1}{3}
\mbox{div}_{\gamma}\nabla 
-2\bar{g}({\bf H}_{\rm N}, {\bf H}_{\rm N})\right)a_{\rm N}^2
=\frac{8\pi Ga^2_{\rm N}}{3}T,
\end{align}
where $\partial_{\bf n}a_{\rm N}^2={\bf n}(a_{\rm N}^2)$  and 
$\partial_{{\bf n}^2}^2a_{\rm N}^2=\partial_{\bf n}(\partial_{\bf n}a_{\rm N}^2)
$.
It is important to recognize that this final form of the second Friedmann equation aligns with the first order equations of GR.
\begin{remark}
    Dealing with the conformal metric $d\bar{s}$, the final version  of the first Friedmann equation, which up to first order matches with the field equations of GR,  acquires the form:
\begin{align}
    \bar{g}({\bf H}_{\rm N}, {\bf H}_{\rm N})a_{\rm N}^2-\frac{1}{3}\mbox{div}_{\gamma}(\nabla a_{\rm N}^2)=\frac{8\pi Ga_{\rm N}^4}{3}\rho. 
\end{align}
  
\end{remark}

Summing up, 
the perturbed Friedmann equations are:
\begin{align}
&\frac{1}{a_{\rm N}^2}\left( \bar{g}({\bf H}_{\rm N}, {\bf H}_{\rm N})-\frac{1}{3}\mbox{div}_{\gamma}\nabla \right)a_{\rm N}^2    =\frac{8\pi Ga_{\rm N}^2} {3}\rho,\\ 
& \frac{1}{a_{\rm N}^2}\left(
\partial^2_{{\bf n}^2}
-\frac{1}{3}
\mbox{div}_{\gamma}\nabla 
-2\bar{g}({\bf H}_{\rm N}, {\bf H}_{\rm N})\right)a_{\rm N}^2
=\frac{8\pi Ga^2_{\rm N}}{3}T.   \end{align}

{In terms of the {\it cosmic time-like vector}  $\partial_{\bf t}=\frac{1}{a_{\rm N}}\partial_{\bf n}$, which is the normal time-like vector with respect to the metric $g$, the Hubble vector rate has the form
${\bf H}_{\rm N}=a_{\rm N}^2H_{\rm N}\partial_{\bf t}$,
where we have introduced the Hubble rate in {\it cosmic time} $H_{\rm N}\equiv 
\frac{\partial_{\bf t} a_{\rm N}}{a_{\rm N}}
$, and the Friedmann equations become:
}
\begin{align}
&\left( \frac{1}{a_{\rm N}^2}{\frak{g}}({\bf H}_{\rm N}, {\bf H}_{\rm N})-\frac{1}{3}\mbox{div}_{\gamma}\nabla \right)a_{\rm N}^2  =\frac{8\pi Ga_{\rm N}^4} {3}\rho,\\    
& \left(
a_{\rm N}\partial_{\bf t} (a_{\rm N}\partial_{\bf t})
-\frac{1}{3}
\mbox{div}_{\gamma}\nabla 
-\frac{2}{a_{\rm N}^2}{\frak{g}}({\bf H}_{\rm N}, {\bf H}_{\rm N})\right)a_{\rm N}^2
=\frac{8\pi Ga^4_{\rm N}}{3}T.  
\end{align}
Combining both expressions, 
 we arrive to the more familiar form of the  Friedmann equations:
\begin{eqnarray}
&& H_{\rm N}^2- \frac{1}{3a_{\rm N}^4}\mbox{div}_{\gamma}(\nabla a_{\rm N}^2)=\frac{8\pi G}{3}\rho,\\
&& \frac{\partial^2_{{\bf t}^2}a_{\rm N}}{a_{\rm N}}+\frac{1}{6a_{\rm N}^4}\mbox{div}_{\gamma}(\nabla a_{\rm N}^2)=-\frac{4\pi G}{3}(\rho+3p).
\end{eqnarray}

{Taking into account the relation
\begin{align}
\mbox{div}_{\gamma}\nabla (a_{\rm N}^2)=2(\gamma(\nabla a_{\rm N}, \nabla a_{\rm N})+a_{\rm N}\mbox{div}_{\gamma}(\nabla a_{\rm N})), 
\end{align}
and disregarding the second order term
$\gamma(\nabla a_{\rm N}, \nabla a_{\rm N})$,
the perturbed Friedmann equations become:
\begin{eqnarray}
&& H_{\rm N}^2- \frac{2}{3a_{\rm N}^3}\mbox{div}_{\gamma}(\nabla a_{\rm N})=\frac{8\pi G}{3}\rho,\\
&& \frac{\partial^2_{{\bf t}^2}a_{\rm N}}{a_{\rm N}}+\frac{1}{3a_{\rm N}^3}\mbox{div}_{\gamma}(\nabla a_{\rm N})=-\frac{4\pi G}{3}(\rho+3p),
\end{eqnarray}
{or introducing the space-like Hubble vector ${\bf H}_{\gamma, \rm N}=\frac{\nabla a_{\rm N}}{a_{\rm N}}$ and taking into account that
$\mbox{div}_{\gamma}(\nabla a_{\rm N})=\gamma(\nabla a_{\rm N},{\bf H}_{\gamma,\rm N})+a_{\rm N}\mbox{div}_{\gamma}
{\bf H}_{\gamma,\rm N}$, at the first order:
\begin{eqnarray}
&& H_{\rm N}^2- \frac{2}{3a_{\rm N}^2}\mbox{div}_{\gamma}{\bf H}_{\gamma,\rm N}=\frac{8\pi G}{3}\rho,\\
&& \frac{\partial^2_{{\bf t}^2}a_{\rm N}}{a_{\rm N}}+\frac{1}{3a_{\rm N}^2}\mbox{div}_{\gamma}{\bf H}_{\gamma,\rm N}=-\frac{4\pi G}{3}(\rho+3p),
\end{eqnarray}
}which together the conservation equation
$\nabla_{\mu}T^{\mu}_{\nu}=0$, that at the first order leads to:
\begin{align}\label{Conservation_Euler}
& \partial_{\bf t}\rho+(\rho+p)[3H_{\rm N}+\mbox{div}_{\gamma}{\bf v}]=0,
\\
& \partial_{\bf t}(a_{\rm N}^2(\rho+p){\bf v})+
 (\rho+p)[
 3H_{\rm N}a_{\rm N}^2{\bf v}-{\bf H}_{\gamma,\rm N} ]
 +\nabla p
=0,\label{Euler_final}
\end{align}
where ${\bf v}\equiv \frac{d{\bf q}}{ds}$, are the dynamical equations, aligning  at the first order of perturbations, with the ones of GR.
To end, two remarks are in order:

\begin{remark}
    The combination of the two Friedmann equations and the conservation equation leads to the constraint:
    \begin{align}
\mbox{div}_{\gamma}(\nabla\partial_{\bf t}a_{\rm N})=
        4\pi Ga_{\rm N}^3(\rho+p)\mbox{div}_{\gamma}{\bf v}.
        \end{align}

In addition, 
the second Friedmann equation can be written as:
\begin{align}
\label{Ricci}\partial_{\bf t}H_{\rm N}+\frac{1}{a_{\rm N}^2}
\mbox{div}_{\gamma}
{\bf H}_{\gamma,\rm N}=-4\pi G(\rho+p),\end{align}
or
\begin{align}
2\partial_{\bf t}H_{\rm N}+3H_{\rm N}^2=-8\pi G p,
\end{align}
and thus, recalling that the first Friedmann equation is a constraint,  one can take as a dynamical equations (\ref{Ricci}) together with (\ref{Conservation_Euler}) and (\ref{Euler_final}).

\end{remark}

}

{
\begin{remark}
    In terms of the four-gradient of a scalar function $f$, i.e., 
$\mbox{grad}_{\frak{g}}(f)=g^{\mu\nu}\partial_{\mu}f\partial_{\nu}$
    and the four-divergence of a four-vector ${\bf w}$, i.e., $\mbox{div}_{\frak{g}} {\bf w}=\frac{1}{\sqrt{-g}}\partial_{\mu}(\sqrt{-g}w^{\mu})$, the second Friedmann equation can be written in the geometric form:
\begin{eqnarray}
\mbox{div}_{\frak{g}}
{\bf H}_{\frak{g},\rm N}
=\frac{4\pi G}{3}T,
\end{eqnarray}
where
\begin{align}
{\bf H}_{\frak{g},\rm N}\equiv\frac{1}{a_{\rm N}}\mbox{grad}_{\frak{g}}(a_{\rm N})=H_{\rm N}\partial_{\bf t}-\frac{1}{a_{\rm N}^2}{{\bf H}}_{\gamma,\rm N},
\end{align}
is the four-Hubble vector.
In fact, introducing the four-velocity ${\bf w}$,
we can write the stress tensor as: 
\begin{eqnarray}
    \frak{T}=(\rho+p){\bf w}\otimes{\bf w}-p\frak{g},
\end{eqnarray}
 and  the dynamical equations acquire the following simple geometric form:
\begin{eqnarray}
\mbox{div}_{\frak{g}}{\bf H}_{\frak{g},\rm N}=\frac{4\pi G}{3}T,\qquad
\mbox{div}_{\frak{g}}\frak{T}=0.
\end{eqnarray}

\end{remark}

}

\section{Conclusions}
\label{sec-conclusion}

\noindent In this article, we have heuristically derived the relativistic Friedmann equations at first-order perturbations, starting from the principles of Newtonian mechanics and the first law of thermodynamics. This approach provides with an intuitive yet rigorous pathway to understand the fundamental equations governing the dynamics of the universe.\\

\noindent This alternative methodology for deriving the Friedmann equations establishes a robust framework for exploring the evolution of the cosmos. By synthesizing the principles of Newtonian mechanics and thermodynamics, we gain valuable insights into the intricate interplay between various physical processes, including the relationships between energy density, pressure, and spacetime geometry. This synthesis not only serves as a pedagogical tool but also deepens our conceptual understanding of the mechanisms driving the cosmic expansion and evolution.\\

\noindent Furthermore, we extended our study to include the derivation of perturbation equations in the Newtonian gauge within the framework of GR. These equations are rooted in the classical conservation laws governing a perfect fluid: the continuity equation, which ensures the conservation of mass, and the Euler equation, which describes the conservation of momentum. By incorporating these classical laws into the relativistic context, we derived perturbative equations that describe the behavior of matter and spacetime under small deviations from homogeneity and isotropy.\\

\noindent Through this comprehensive approach, we bridge the gap between classical mechanics and GR, linking the intuitive foundations of Newtonian physics with the profound insights of modern cosmology. This work not only enhances our theoretical understanding of the universe's dynamic evolution but also provides a unified perspective on how classical and relativistic descriptions converge in the study of cosmological phenomena.

\begin{acknowledgments}
JdH is supported by the Spanish grant 
PID2021-123903NB-I00
funded by MCIN/AEI/10.13039/501100011033 and by ``ERDF A way of making Europe''. EE  has been partially supported by MICINN (Spain), project 2024AEP171,
of the Spanish State Research Agency program AEI/10.13039/501100011033,
by AGAUR project 2021-SGR-00171, and by the program Unidad de Excelencia
María de Maeztu CEX2020-001058-M. SP has been supported by the Department of Science and Technology (DST), Govt. of India under the Scheme   ``Fund for Improvement of S\&T Infrastructure (FIST)'' (File No. SR/FST/MS-I/2019/41).  
\end{acknowledgments}

\section*{APPENDIX A: The Newtonian gauge and the Principle of Equivalence}

\noindent The objective of this appendix is to provide a justification for the Newtonian gauge without relying on the field equations of GR, as discussed in page 182 of \cite{Weinberg:1972kfs}, where S. Weinberg states: {\it ``The general kinematic framework provided by the Principle of Equivalence rests on a much firmer foundation than do Einstein's field equations. Indeed, in Chapters 3 and through 5 we were led almost inevitably from the equality of gravitation and inertial mass to the fill formalism of tensor analysis and general covariance, whereas in contrast the derivation of Einstein's equations in Chapter 7 contained a strong element of guesswork, and in any case there might exist a long range scalar field, like that of Brans and Dicke, that would alter the field equations. It is very useful to test general relativity by assuming that the usual rules for the motion of particles and photons in a given metric still apply, but that the metric may be different from that calculated from the Einstein equations''}. \\

\noindent Following Weinberg's approach, we draw upon Einstein's foundational ideas for constructing a theory of gravitation consistent with his theory of special relativity, particularly the initial formulation of his Principle of Equivalence. Here, we reproduce the version quoted in \cite{pauli1981theory}: {\it ``In the Newtonian theory a system,  a homogeneous gravitational field is completely equivalent to a uniform accelerated reference system, from a mechanical point of view'',}
or in Einsteins's words~\cite{Einstein1}: {\it ``A falling man is accelerated. Then what he
feels and judges is happening in the
accelerated frame of reference. I decided to extend the theory of relativity to
the reference frame with acceleration.
I felt that in doing so I could solve the
problem of gravity at the same time. A
falling man does not feel his weight
because in his reference frame there is
a new gravitational field which cancels the gravitational field due to the Earth.
In the accelerated frame of reference,
we need a new gravitational field.''}\\

\noindent It is common lore that Einstein got this idea (reportedly, ``the happiest of his life'') {\bf in} an ordinary day, while he was sitting at his table in Bern's patent office. He tried to guess what would happen if, at that very moment, he fell from the roof of his house, in an upright position. If he had an object on the palm of his hand, the object would not fall down to his feet when he would withdraw his hand, he concluded.
It is important to recognize that when Einstein formulated his Principle of Equivalence, it was not yet known that freely falling particles follow geodesics in curved spacetime. 
No wonder, since the formulation of general relativity in terms of differential geometry, concepts were not fully in place at that time, yet.
Therefore, the initial version of the Equivalence Principle must be understood within the framework of Newtonian gravity.\\

\noindent Following this perspective, we begin with Newton's second law applied to a uniform gravitational field
in the 
$z$-direction. In this scenario, a stationary observer examines the motion of a freely falling particle with mass 
$m$.  We adopt the
``Weak Equivalence Principle'' \cite{Weinberg:1972kfs}, i.e., the
standard assumption which states that the gravitational mass and inertial mass are equivalent, a cornerstone of Einstein's Equivalence Principle. The observer uses Newton's second law to describe the particle's motion:
\begin{eqnarray}
    ma_c=-mg\equiv F_g,
\end{eqnarray}
where $a_c$ is the acceleration of the particle, and $F_g$
is the gravitational force acting on it. 
 Rearranging, we can express this as:
\begin{eqnarray}
    F_i+F_g=0,
\end{eqnarray}
where $F_i\equiv -ma_c$ is the inertial force. This equation implies that, from the perspective of a free-falling mass, the net force acting on it is zero. In other words, the mass does not experience its weight, and, from its point of view, it belongs to an inertial reference system. This observation embodies Einstein's Principle of Equivalence, 
i.e., the  ``Weak Principle of Equivalence'', 
introduced in 1907 \cite{Einstein1953-EINTMO-8}.\\

\noindent Nevertheless, there is a more traditional infinitesimal form of this principle
proposed by Pauli \cite{pauli1981theory} (see also \cite{NORTON1985203} for a detailed discussion of Einstein's beliefs regarding the Equivalence Principle). In some sense, this principle does not fully align with Mach's Principle, as discussed in page 87 of \cite{Weinberg:1972kfs}. From our perspective, geometrically, this principle simply states the well-known fact that the tangent space of a Lorentz manifold 
 $\mathcal M$— 
 is the Minkowski space. 
 Specifically, given an infinitesimal chart
$(U, \varphi)$, where $\varphi:U\rightarrow \R^4$, 
and considering a point 
$P$ on a geodesic, this geodesic can be  approximated --at zero order-- in its tangent space at 
$P$ (which corresponds to Minkowski space). To perform calculus in this tangent space, one must return to coordinates,
 i.e., to $\R^4$.
 The simplest computations occur when one chooses a chart where the geodesics appear as straight lines (co-ordinates where the Christoffel symbols vanishes at point $P$,  and for points close to $P$ one takes the zero order approximation of the symbols), which, in physical terms, corresponds to locally inertial frames.\\

\noindent Thus, we arrive at the following result: in the tangent space, the geodesics of the manifold can be locally approximated by straight lines when an appropriate chart, i.e., a locally inertial frame, is chosen. By itself, this principle may seem trivial. However, when combined with the principle of covariance, it provides a framework for incorporating the physics of special relativity, such as electromagnetic dynamics, into the Lorentz manifold. 
This is accomplished by the straightforward rule of replacing ordinary derivatives with their covariant generalizations and standard products in Minkowski space with those defined on the manifold.\\

\noindent It is important to note, however, that  higher-order terms do not appear in the zero-order approximation. Consequently, the principle of covariance only transfers zero-order effects to the manifold. Therefore, constructing a theory based on this approach may fail to capture all interactions between gravity and other fields if such interactions involve terms of order greater than zero (see, for instance \cite{Wilczek,Elizalde:2023gbs}, to delve deeper into this line of thought). \\
 
\noindent The remaining challenge is to determine the geometry of the Lorentzian manifold, which, following Mach's principle, can be established by relating the geometry of the manifold to the matter content of the universe through specific field equations. This monumental task was completed by Einstein in 1915, where he connected the Ricci tensor and scalar to the stress-energy tensor.\\

\noindent At this point, it is important to note that the Ricci tensor and scalar are invariant under diffeomorphisms. This leads to the Hole Argument (see for instance \cite{deHaro:2023bsj}), which continues to provoke discussions in the philosophical community \cite{Stachel,Earman-Norton,Pooley:2020mar,Pooley:2021hct}. The physical rejection of the Hole Argument entails a completely different perspective on space-time, challenging the substantivalist notion—a fundamental concept in Newtonian doctrine and even in special relativity. According to Einstein, this substantivalist view must be replaced, in line with Descartes' and Leibniz's relationalism, by a framework based on ``point coincidences''. \\

\noindent Building on the weak principle, Einstein sought to extend his special theory of relativity to include accelerated reference systems. To achieve this, he considered two reference systems: an inertial frame
$K$ with co-ordinates $(T,X_1,X_2,X_3)$, and another frame $K'$ with co-ordinates $(t,x_1,x_2,x_3)$, which is uniformly accelerated with a constant acceleration
 $g$ in the $X_3$-direction relative to  $K$.  
On the other hand, the Principle of Equivalence tells us that the frame $K'$ is equivalent to a frame at rest with a uniform gravitational field $\Phi_{\rm N}(x_3)=gx_3$, and therefore the free-falling mass does not experience its weight, meaning that its inertial  and gravitational forces compensates, that is:
\begin{eqnarray}\label{inertia}
    -ma_c-mg=0\Longleftrightarrow a_c=-g.
\end{eqnarray}
The transformation of co-ordinates is: 
\begin{eqnarray}
\left\{\begin{array}{ccc}
T=\left(\frac{1}{g}+x_3\right)
  \sinh(gt)\\
\bigskip \\
X_3=-\frac{1}{g}+\left(\frac{1}{g}+x_3\right)\cosh(gt), 
\end{array}\right.
\end{eqnarray}
leading to the line element
\begin{eqnarray}
    ds^2=(1+gx_3)^2dt^2-(dx_1^2+dx_2^2+dx_3^2).
\end{eqnarray}
In the frame $K'$, considering 
the action
$S=-m\int ds$ and making its variation we find the dynamical equations for a free falling mass
$\delta S=0$,  we find:
\begin{eqnarray}\label{dyn_eqs}
   \frac{ds}{dt}\left[\frac{d}{dt}
\left(\frac{dx_3}{ds}\right)\right]
    =-g(1+gx_3).
\end{eqnarray}
The question arises: what is the value of the acceleration? At first glance, it seems that one must select the acceleration relative to an observer at rest in the frame 
$K'$,
obtaining $a_c=\frac{d^2x_3}{dt^2}$.
In this scenario, the equations (\ref{dyn_eqs}) and (\ref{inertia}) coincide only under the approximation:
\begin{eqnarray}
\left|\frac{dx_3}{dt}\right|\ll |1+gx_3|\qquad \mbox{and} \qquad |gx_3|\ll 1.
\end{eqnarray}
Hence, the Principle of Equivalence  holds only approximately.
At this point, we seek the metric that reproduces Newton's second law.
To achieve this, we consider the static line element:
\begin{eqnarray}
    ds^2=
    A(x_3)dt^2-(dx_1^2+dx_2^2+B(x_3)dx_3^2),
\end{eqnarray}
and from 
$\delta S=0$, we find:
\begin{align}
    m\frac{d^2x_3}{ds^2}=-\frac{m}{2A(x_3)B(x_3)}\left[
\partial_{x_3}A+\partial_{x_3}(AB)\left(\frac{dx_3}{ds}\right)^2 \right].
\end{align}
To obtain the Newtonian equation, we must require that the velocity on the right-hand side vanishes. This condition is satisfied when:
$A(z)B(z)=C$, being $C$ a constant, which be taken equal to $1$ rescaling the co-ordinates.
Therefore,  we have:
\begin{eqnarray}
    -m\frac{d^2x_3}{ds^2}-\frac{m}{2}\partial_{x_3}A=0,
\end{eqnarray}
and comparing with (\ref{inertia}) we find that
\begin{eqnarray}
a_c=\frac{d^2x_3}{ds^2}\qquad \mbox{and}\qquad 
A(x_3)=b+2\Phi_{\rm N}(x_3),
    \end{eqnarray}
where $b$ is a constant that must be equal to $1$ to recover the Minkowski metric when the acceleration vanishes.\\

\noindent The conclusion is that, in the case of a homogeneous gravitational field $\Phi_{\rm N}(x_3)=gx_3$, the dynamics of a freely falling mass are governed by Newton's second law when the acceleration 
$a_c$ is the proper acceleration of the free-falling particle. Furthermore, we can assert that the mass does not experience its weight because the proper inertial force precisely compensates for the Newtonian gravitational force.
We now extend this result to a Newtonian radial potential 
 $\Phi_{\rm N}(r)$ 
 that approaches zero at infinity and consider a test mass undergoing radial free fall
  (see also
\cite{2019EL....12549001F}). In this scenario, Newton's equation is expressed as,
\begin{eqnarray}
  m\frac{d^2{\bf x}}{ds^2}=-m\partial_r \Phi_{\rm N}\frac{{\bf x}}{r}
  \Longrightarrow
    \frac{d^2r}{ds^2}=-\partial_r \Phi_{\rm N},\end{eqnarray}
where ${\bf x}=(x_1,x_2,x_3)$ and $r=|{\bf x}|$. Now, taking the line element
\begin{eqnarray}\label{Hilbert_metric} ds^2=A(r)dt^2 - B(r)dr^2 - C(r)r^2d\Omega, \end{eqnarray}
where $d\Omega=d\theta^2+\sin^2\theta d\varphi^2$,
and making the variation of the action we arrive at,
\begin{eqnarray}
    -\frac{d^2r}{ds^2}-\frac{1}{2}\partial_rA=0\quad \mbox{provided by} \quad A(r)B(r)=1,
\end{eqnarray}
which, by imposing that the metric approaches the Minkowski metric at infinity, we find,
\begin{align}
A(r)=1+2\Phi_{\rm N}(r), \quad \mbox{and}
    \quad B(r)=(1+2\Phi_{\rm N}(r))^{-1}.
    \end{align}

\noindent In order to determine the value of $C(r)$, 
one may consider the motion in the plane $\theta=\pi/2$. 
In a central potential, the angular momentum
 ${\bf M}\equiv{\bf x}\times \frac{d{\bf x}}{ds}= r^2\frac{d\varphi}{ds}$ remains conserved.
  Furthermore, the variation of the action with respect to 
  $\varphi$ leads to the conservation of $C(r)r^2\frac{d\varphi}{ds}$. From this, it follows that $C(r)$  must be constant, and, in fact, it is equal to 
$1$ when the Minkowski metric is imposed at infinity.   
Let us note that, when one considers the particular case of a point particle with mass $M$ situated at the origin of coordinates, the potential is given by $\Phi_{\rm N}(r) = -\frac{MG}{r}$, resulting in the well-known
Hilbert-Droste  \cite{Hilbert,Droste} version of the original
Schwarzschild metric
\cite{Schwarzschild:1916uq}
\begin{eqnarray}
    A(r)=1-\frac{2MG}{r},~\mbox{and}~  B(r)=\left(1-\frac{2MG}{r}\right)^{-1}.\end{eqnarray}

Now, with the use of the harmonic co-ordinates given by \cite{Weinberg:1972kfs}

\begin{eqnarray}\label{dark}
\left\{\begin{array}{ccc}
X_1=R\sin\theta\cos\varphi,\quad
X_2=R\sin\theta \sin\varphi,\\
\bigskip \\
X_3=R\cos\theta,\qquad t=t, 
\end{array}\right.
\end{eqnarray}
with $R=r-MG$, the Schwarzschild  metric becomes,
\begin{eqnarray}
ds^2=\left(\frac{1-MG/R}{1+MG/R}\right)dt^2-
\left(1+\frac{MG}{R}\right)^2 d{\bf X}^2 \nonumber\\-
\left(\frac{1+MG/R}{1-MG/R}\right)
    \frac{M^2G^2}{R^4}({\bf X}.d{\bf X})^2,
    \end{eqnarray}
which for $MG\ll R$, and retaining only linear terms on $MG/R$, leads to
\begin{align}
ds^2 = & \left(1-\frac{2MG}{R}\right)dt^2-
\left(1+\frac{2MG}{R}\right) d{\bf X}^2\nonumber \\
\implies &
ds^2=\left(1-\frac{2MG}{r}\right)dt^2-
\left(1+\frac{2MG}{r}\right) d{ r}^2
-r^2d\Omega,
    \end{align}
which is the Newtonian gauge for a point particle, and also provides the same perihelion precession of Mercury as the Schwarzschild metric, because both metrics, for $MG/r\ll1$,  lead to the same  key equation
\cite{Weinberg:1972kfs, Magnan:2007uw}:
\begin{eqnarray}
\left(\frac{d\varphi}{du}\right)^2=\frac{1+2MG(u+u_++u_-)}{(u-u_+)(u_--u)},
    \end{eqnarray}
where $u=1/r$, $u_+=1/r_+$ and $r_-=1/r_-$, being $r_-$ is the perihelion distance and $r_+$ the aphelion distance. 
Therefore, it is natural to assume that this gauge also holds in the week limit $|\Phi_{\rm N}({\bf x})|\ll 1$. 
{More precisely, 
denoting by $\dot{\bf x}$ the derivative with respect the proper time 
and considering the metric
\begin{eqnarray}\label{metric}
ds^2=(1+2\Phi_{\rm N}({\bf x}))dt^2-\frac{1}{
    1+2\Phi_{\rm N}({\bf x})    }d{\bf x}^2,
\end{eqnarray}
the variation of the action leads to the dynamical equation:
\begin{eqnarray}
\ddot{\bf x}=
   -\nabla \Phi_{\rm N}-
\frac{2}{
    1+2\Phi_{\rm N}({\bf x})}|\dot{\bf x}|^2\nabla_{\perp}\Phi_{\rm N},    
\end{eqnarray}
where $\nabla_{\perp} \Phi_{\rm N}\equiv[\nabla-\frac{\dot{\bf x}}{  |\dot{\bf x}|^2\   }(\dot{\bf x}.\nabla)]\Phi_{\rm N}$ is the orthogonal projection of $\nabla \Phi_{\rm N}$ onto  
$\dot{\bf x}$. 
For small velocities, it
 coincides with the Newton's. Furthermore,  
   the static  metric (\ref{metric}) exactly aligns with the Newton's law when the movement is in the direction of the gradient of the potential, which happens for homogeneous potentials
 $\Phi_{\rm N}({\bf x})=kx_i$ when the movement is in the direction of the  $x_i$-axis, and for radial movements when the  potentials  are of the form $\Phi_{\rm N}({\bf x})=f(r)$, because in that case
 $\nabla \Phi_{\rm N}(|{\bf x}|)=\frac{df(r)}{dr}\frac{\bf x}{|\bf x|}$ and for a radial trajectory $\dot{\bf x}=\dot{r}\frac{{\bf x}}{|{\bf x}|}$.

\bibliography{references}

\end{document}